# Security Parameter Analysis of the LINEture Post-Quantum Digital Signature Scheme

Yevgen Kotukh[2[0000-0003-4997-620X]] and Gennady Khalimov[1[0000-0002-2054-9186]]

[1] Kharkiv National University of Radio electronics, Kharkiv, Ukraine
[2] Yevhenii Bereznyak Military Academy, Kyiv Ukraine
yevgenkotukh@gmail.com

***Abstract.*** *This paper presents a comprehensive cryptographic analysis of the security parameters of the LINEture post-quantum digital signature scheme, which is constructed using matrix algebra over elementary abelian 2-groups. We investigate the influence of three principal parameters: the word size m (exhibiting quadratic impact), the vector dimension l, and the number of submatrices in the session key q (exhibiting linear impact) on cryptographic strength. Our analysis reveals a dualistic nature of the parameter l: according to the original authors' analysis, it does not affect resistance to guessing attacks; however, a deeper examination of the verification mechanism demonstrates that l establishes a "verification barrier" of l·m bits. We establish the threshold relationship l < (q−1)·m, below which parameter l becomes security-critical. The optimal selection rule l ≈ (q−1)·m is proposed for maximum cryptographic efficiency. Comparative analysis with NIST PQC standards and practical parameter recommendations are provided.*

## I. INTRODUCTION

The advancement of quantum computing poses a fundamental threat to existing cryptographic systems predicated on the computational hardness of integer factorization and discrete logarithm problems. Shor's algorithm enables efficient solution of these problems on a quantum computer, thereby necessitating the development of alternative cryptographic primitives resistant to quantum attacks [1].

The National Institute of Standards and Technology (NIST) concluded its Post-Quantum Cryptography Standardization process in 2024, selecting lattice-based schemes (CRYSTALS-Dilithium, Falcon) and hash-based signatures (SPHINCS+) as the primary standards [2]. Nevertheless, the exploration of alternative constructions remains pertinent owing to the imperative for cryptographic diversity and the potential for more efficient solutions in resource-constrained environments.

The LINEture cryptosystem represents a novel post-quantum digital signature scheme founded upon linear matrix algebra over elementary abelian 2-groups of order $2m$. The system employs a zero-knowledge proof mechanism within the Fiat-Shamir paradigm [3]. A distinguishing advantage of this system is its key compactness: approximately 300-500 bytes compared to 2-3 KB for Dilithium.

Despite the theoretical appeal of this construction, the rigorous justification of security parameters for LINEture remains insufficiently investigated. In particular, the role of parameter l-the dimension of the substitution vector-in ensuring cryptographic strength is ambiguous and warrants deeper analysis.

The objective of this work is to conduct a comprehensive analysis of the influence of parameters (m, l, q) on the security of the LINEture cryptosystem, to identify latent dependencies, and to formulate recommendations for optimal parameter selection.

## II. MATHEMATICAL FOUNDATIONS

*A. Parameter Structure*

The LINEture cryptosystem is defined by three principal parameters, enumerated in Table I.

TABLE I. Primary Cryptosystem Parameters

| Parameter | Values | Function |
|---|---|---|
| $m$ | 8 or 16 bits | Word size (substitution element) |
| $l$ | 8 or 16 | Vector dimension (number of m-bit words) |
| $q$ | 3 or 4 | Number of submatrices in session key K |

*B. Session Key Structure*

The session key K is constructed as a concatenation of q submatrices, each of dimension m×m:

$$K = [H_1 \mid H_2 \mid R_1 \mid R_2 \mid ... \mid R_{q-2}] \quad (1)$$

where $H_1$ and $H_2$ are submatrices deterministically derived from message hashes $h(r_1, msg)$ and $h(r_2, msg)$, and $R_i$ are random secret submatrices. The cardinality of the secret key space is:

$$|K_{secret}| = 2^{m^2 \cdot (q-1)} \quad (2)$$

*C. Shared Secret and Substitution Vector*

The shared secret $S = (S_1, S_2, ..., S_l)$ comprises a vector of $l$ bijective substitutions $S_i: \{0,1\}^m \to \{0,1\}^m$. A critical property is that the secret is computed for each of the $l$ components **independently**, yet utilizing a **common** session key K:

$$S_i = f(B_i, K), \quad i = 1, ..., l \quad (3)$$

*D. Cryptographic Primitive Sizes*

Based on structural analysis of the cryptosystem, the following size relationships have been established:
- Public key: $pk = 6 \cdot l \cdot m^2/8$ bytes
- Secret key: $sk = l \cdot m^2/8$ + auxiliary parameters
- Signature: $sign = l \cdot m/8 + 2(q-1) \cdot m^2/8 + 64$ bytes

## III. SECURITY ANALYSIS

*A. Session Key Guessing Attack*

According to the original security analysis presented by the cryptosystem's authors, the principal attack vector is session key guessing. The security level is characterized by:

$$S_{guessing} = 2 \cdot (q-1) \cdot m^2 \text{ bits} \quad (4)$$

**Key observation by the authors:** parameter $l$ is **absent** from equation (4). This is corroborated by the security table from the original work (Table II).

TABLE II. Security Levels

| (m, l, q) | $S_{guessing}$ | $S_{collision}$ |
|---|---|---|
| (8, 8, 3) | 256 bits | ~224 bits |
| (8, 16, 3) | 256 bits | ~224 bits |
| (16, 8, 3) | 1024 bits | ~896 bits |
| (16, 16, 3) | 1024 bits | ~896 bits |

As evidenced by Table II, configurations (8,8,3) and (8,16,3) exhibit **identical** guessing security (256 bits), notwithstanding different values of l.

*B. Collision Attack*

The collision attack is independent of parameters l and q, being determined solely by the word size m:

$$S_{collision} \approx 3.5 \cdot m^2 \text{ bits} \quad (5)$$

For m=8, this yields approximately 224 bits, which constitutes the **limiting factor** for small values of m regardless of increases in q.

*C. Parameter Influence per Authors' Analysis*

According to the original analysis, the influence of parameters on security is characterized as follows:
- *m* - **critical** (quadratic influence, $m^2$): the fundamental unit of security
- *q* - **significant** (linear influence): each increment of q adds $2m^2$ bits
- *l* - **auxiliary**: determines only hash size ($l \times m$ bits) and key dimensions

*A. Signature Verification Mechanism*

A detailed analysis of the verification procedure reveals a latent role for parameter l. During verification, the satisfaction of **l independent equations** is verified:

$$S_i(z_i) = h_i, \quad \forall i = 1, ..., l \quad (6)$$

In a guessing attack, the adversary generates a candidate K' and computes $S' = (S'_1, ..., S'_l)$. For a successful attack, **all l equations** must be simultaneously satisfied.

*B. Probabilistic Analysis*

If K' is incorrectly guessed, then with high probability $S'_i \neq S_i$. The probability of fortuitous coincidence for a single equation is:

$$P(S'_i(z_i) = h_i \mid S'_i \neq S_i) = 2^{-m} \quad (7)$$

For *all l equations* simultaneously:

$$P_{success} = (2^{-m})^l = 2^{-l \cdot m} \qquad (8)$$

**Fundamental conclusion:** parameter l establishes an additional "verification barrier" of **l·m bits**, which constrains the space of viable attacks by a factor of $2^{l \cdot m}$.

*C. Linear-Algebraic Analysis*

For identity proof, two session keys $K_1$ and $K_2$ must yield an identical shared secret:

$$A \cdot K_1 \cdot B = A \cdot K_2 \cdot B = S \qquad (9)$$

This condition engenders a system of l·m linear equations over GF(2). The effective attack space becomes:

$$2^{2(q-1)m^2 - l \cdot m} \qquad (10)$$

*D. Threshold Value for Parameter l*

Parameter l influences security when l·m < (q−1)·m², that is:

$$l < (q-1) \cdot m \qquad (11)$$

This defines the threshold value for parameter l (Table III).

TABLE III. Threshold Values for Parameter l

| m | q | Threshold l | l=8 Status |
|---|---|---|---|
| 8 | 3 | 16 | Critical |
| 8 | 4 | 24 | Critical |
| 16 | 3 | 32 | Critical |

To visualize the influence of parameters on security and cryptosystem efficiency, a comprehensive graphical analysis has been constructed (Figs. 1 and 2).

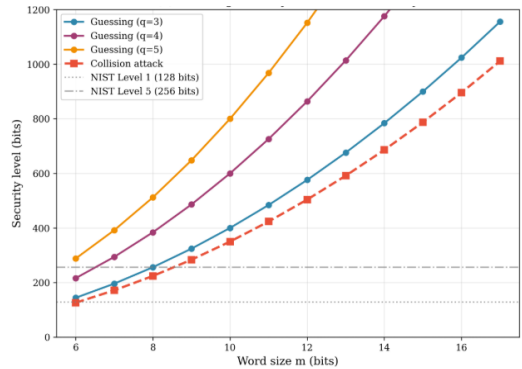

Fig. 1 – Guessing Security vs. Collision Security

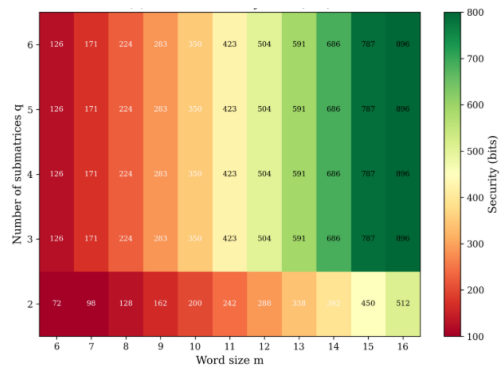

Fig. 2 – Security heatmap of minimal security level

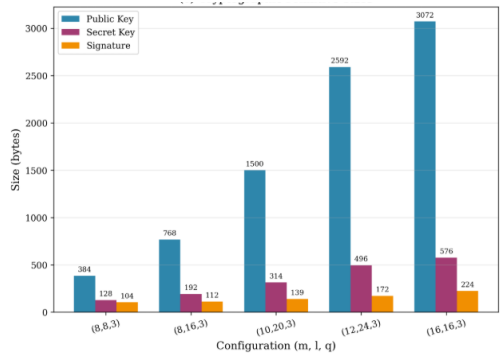

Fig. 3 – SK, PK and Digital Signature sizes

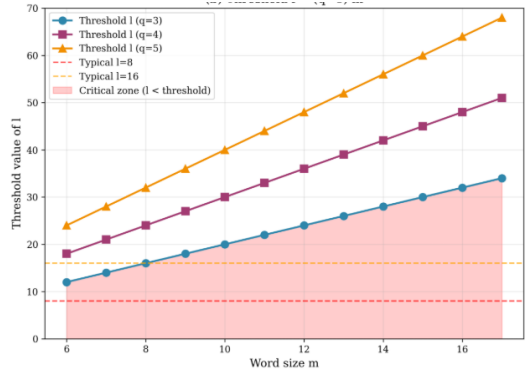

Fig. 4 – Threshold l=(q-1)m

# IV. OPTIMAL PARAMETER SELECTION

*A. Synthesis of Results*

Integrating the authors' analysis with our deep analysis of the verification barrier, we formulate a refined security model:

$$S_{effective} = min(2(q-1)m^2,\ 3.5m^2,\ l \cdot m + (q-1)m^2/2) \quad (12)$$

where the third term accounts for the verification barrier when $l < (q-1) \cdot m$.

*B. Rules for Optimal Selection*

Based on the analysis conducted, we propose the following selection rules:

**Rule 1 (optimal l):** $l_{opt} = (q-1) \cdot m$. Values below this threshold result in security degradation due to an insufficient verification barrier; values above yield size increases without security gains.

**Rule 2 (m selection):** For resistance to collision attacks, $m \geq 10$ is recommended. At $m=8$, collision resistance is limited to approximately 224 bits regardless of other parameters.

**Rule 3 (q scaling):** Increasing q is effective for enhancing guessing resistance ($+2m^2$ bits per unit increment of q) without substantial key size growth.

TABLE IV. Recommended Configurations

| Level | (m, l, q) | Security | Size | Efficiency |
|---|---|---|---|---|
| Basic | (8, 16, 3) | 224 bits | ~800 B | 0.28 |
| Standard | (10, 20, 3) | 350 bits | ~1.2 KB | 0.29 |
| High | (12, 24, 3) | 504 bits | ~1.8 KB | 0.28 |

*C. Comparison with NIST PQC Standards*

TABLE V. Comparison with NIST PQC Standards

| Scheme | Security | Signature | PK |
|---|---|---|---|
| LINEture (10,20,3) | ~350 bits | ~450 B | ~750 B |
| Dilithium-II | ~128 bits | ~2420 B | ~1312 B |
| Falcon-512 | ~128 bits | ~666 B | ~897 B |

LINEture demonstrates a substantial advantage in signature compactness (3-5× smaller) while providing a higher security level. However, it should be noted that NIST standards have undergone rigorous cryptanalysis, whereas LINEture requires additional independent scrutiny.

## V. DISCUSSION

*A. Dualistic Nature of Parameter l*

From the perspective of direct guessing attacks (equation 4), l is indeed absent from the security expression - this is corroborated by the authors' tables. However, a deeper examination of the verification mechanism (equation 8) reveals a latent "verification barrier."

The reconciliation of these perspectives lies in understanding that when $l \geq (q-1) \cdot m$, the verification barrier becomes redundant and does not affect effective security. However, for typical values l=8 with m=8 and q=3, the condition $l < (q-1) \cdot m = 16$ is satisfied, rendering parameter l critical to security.

*B. Critical Observations on the Cryptosystem*

Notwithstanding its attractive characteristics, LINEture exhibits several significant limitations:

1. **Absence of formal security proofs**-the system lacks reduction to recognized computationally hard problems.
2. **Linear algebraic structure**-potentially vulnerable to algebraic attacks not addressed in the original work.
3. **Quantum resistance is not rigorously established**-the claimed post-quantum security requires formal demonstration.
4. **Absence of independent cryptanalysis**-the system has not undergone public competition analogous to NIST PQC.

## VI. CONCLUSIONS

This paper has presented a comprehensive cryptographic analysis of the security parameters of the LINEture post-quantum digital signature cryptosystem. The principal findings are as follows:

1. The quadratic influence of parameter m and linear influence of q on guessing attack resistance has been established according to the formula $S = 2(q-1) m^2$ bits.
2. The dualistic nature of parameter l has been identified: per the authors' assessment it does not affect security; however, deep analysis of verification reveals the formation of a "verification barrier" of $l \cdot m$ bits.
3. The threshold relationship $l < (q-1) \cdot m$ has been established, below which parameter l becomes security-critical.
4. The optimal selection rule $l_{opt} = (q-1) \cdot m$ has been formulated for maximum cryptographic efficiency.
5. Recommendations for three security levels have been developed, accounting for the security-size trade-off.

We consider the following directions for further research. First, a formal security proof in the random oracle model, the second one is an analysis of resistance to quantum algebraic attacks and comparative analysis of software and hardware implementation efficiency.

## REFERENCES

[1] Khalimov, G., Kotukh, Y., Kolisnyk, M., Khalimova, S., Sievierinov, O., & Volkov, O. (2024, November). SIGNLINE: Digital signature scheme based on linear equations cryptosystem. In *2024 4th International Conference on Electrical, Computer, Communications and Mechatronics Engineering (ICECCME)* (pp. 1-9). IEEE.

[2] Khalimov, G., Kotukh, Y., Kolisnyk, M., Khalimova, S., Sievierinov, O., & Korobchynskyi, M. (2025, March). Digital Signature Scheme Based on Linear Equations. In *Future of Information and Communication Conference* (pp. 711-728). Cham: Springer Nature Switzerland.

[3] Gennady Khalimov, Yevgen Kotukh, Maksym Kolisnyk, Svitlana Khalimova, Oleksandr Sievierinov. LINE: Cryptosystem based on linear equations for logarithmic signatures. https://eprint.iacr.org/2024/697.pdf, Paper 2024/697.

[4] Khalimov, G., Kotukh, Y., & Khalimova, S. (2020, December). Encryption scheme based on the automorphism group of the Ree function field. In *2020 7th International Conference on Internet of Things: Systems, Management and Security (IOTSMS)* (pp. 1-8). IEEE.

[5] Khalimov, G., Sun, X., Kotukh, Y., & Dzhura, I. New Key Encapsulation Method. In *Proceedings of IEMTRONICS 2025: International IoT, Electronics and Mechatronics Conference, Volume 2* (p. 55). Springer Nature.

[6] Khalimov, G., Kotukh, Y., Shonia, O., Didmanidze, I., Sievierinov, O., & Khalimova, S. (2020, October). Encryption scheme based on the automorphism group of the Suzuki function field. In *2020 IEEE International Conference on Problems of Infocommunications. Science and Technology (PIC S&T)* (pp. 383-387). IEEE.

[7] G. Khalimov, Y. Kotukh, S. Khalimova, M. Korobchynskyi, V. Podlipaiev, O. Zaitsev. Cryptanalysis of a LINE cryptosystem. 2025 5th International Conference on Electrical, Computer, Communications and Mechatronics Engineering (ICECCME), Zanzibar, Tanzania, United Republic of, 2025, pp. 1-7, doi: 10.1109/ICECCME64568.2025.11277439.

[8] Khalimov, G., Kotukh, Y., Didmanidze, I., & Khalimova, S. (2021, July). Encryption scheme based on small Ree groups. In *Proceedings of the 2021 7th International Conference on Computer Technology Applications* (pp. 33-37).

[9] Khalimov, G., & Kotukh, Y. (2025). Secured Encryption scheme based on the Ree groups. *arXiv preprint arXiv:2504.17919*.

[10] Khalimov, G., & Kotukh, Y. (2025). MST3 Encryption improvement with three-parameter group of Hermitian function field. *arXiv preprint arXiv:2504.15391*.

[11] Khalimov, G., & Kotukh, Y. (2025). Improved MST3 Encryption scheme based on small Ree groups. *arXiv preprint arXiv:2504.10947*.

[12] Khalimov, G., & Kotukh, Y. (2025). Advanced MST3 Encryption scheme based on generalized Suzuki 2-groups. *arXiv preprint arXiv:2504.11804*.

[13] Khalimov, G., & Kotukh, Y. (2025). Cryptographic Strengthening of MST3 cryptosystem via Automorphism Group of Suzuki Function Fields. *arXiv preprint arXiv:2504.07318*.

[14] G. Khalimov, Y. Kotukh, Yu. Serhiychuk, O. Marukhnenko. "Analysis of the implementation complexity of cryptosystem based on Suzuki group" Journal of Telecommunications and Radio Engineering, Volume 78, Issue 5, 2019, pp. 419-427. DOI: 10.1615/TelecomRadEng.v78.i5.40 21.

[15] Y. Kotukh. "On universal hashing application to implementation of the schemes provably resistant authentication in the telecommunication systems" Journal of Telecommunications and Radio Engineering, Volume 75, Issue 7, 2016, pp. 595-605. DOI: 10.1615/TelecomRadEng.v75.i7.30

[16] Kotukh, Y. V., Khalimov, G. Z., & Dzhura, I. Y. (2025). Криптографічна конкурентоспроможність криптосистем на основі некомутативних груп. *Radiotekhnika*, (221), 72-82.

[17] Котух, Є., Марухненко, О., Халімов, Г., & Коробчинський, М. (2023). Розробка методики випробувань бібліотеки криптографічних перетворювань на прикладі криптосистеми MST3 на основі узагальнених Сузукі 2-груп. *Електронне фахове наукове видання «Кібербезпека: освіта, наука, техніка»*, *2*(22), 113-121.

[18] Kotukh, E. ., Severinov, O. ., Vlasov, A. ., Kozina, L. ., Tenytska, A. ., & Zarudna , E. . (2021). Methods of construction and properties of logarithmic signatures . *Radiotekhnika*, *2*(205), 94–99. https://doi.org/10.30837/rt.2021.2.205.09

[19] Kotukh, Y., & Khalimov, H. (2024). Advantages of Logarithmic Signatures in the Implementation of Crypto Primitives. *Challenges and Issues of Modern Science*, *2*, 296-299. https://cims.fti.dp.ua/j/article/view/119

[20] Kotukh, E., Severinov, O., Vlasov, A., Kozina, L., Tenytska, A., & Zarudna, E. (2021). Methods of construction and properties of logarithmic signatures. *Radiotekhnika*, *2*(205), 94-99.